\documentclass[fleqn,12pt,twoside]{article}
\usepackage[headings]{espcrc1}
\readRCS
$Id: espcrc1.tex,v 1.2 2004/02/24 11:22:11 spepping Exp $
\usepackage{graphicx}
\usepackage[figuresright]{rotating}
\usepackage{amssymb}
\usepackage{sublabel}
\usepackage{graphicx}
\usepackage{epsf}
\usepackage{wrapfig}
\usepackage{epsfig}
\def\Vec#1{\mbox{\boldmath $#1$}}

\def\n{\noindent}
\def\beq{\begin{equation}}
\def\eeq{\end{equation}}

\def\beqy{\begin{eqnarray}}
\def\eeqy{\end{eqnarray}}

\newcommand{\AmS}{{\protect\the\textfont2
  A\kern-.1667em\lower.5ex\hbox{M}\kern-.125emS}}
\title{Slow proton production in semi-inclusive DIS off nuclei: the
role of final state interaction}
    \author{M. Alvioli
  \address{Department of Physics, University of Perugia and
  Istituto Nazionale di Fisica Nucleare, Sezione di Perugia,
  Via A. Pascoli, I-06123, Perugia, Italy},{C. Ciofi degli Atti\addressmark} and V. Palli\addressmark}

\runtitle{Slow proton production in SIDIS}

\begin{document}
\maketitle
\begin{abstract}
\n The effects of the final state interaction on the production of slow protons
in semi-inclusive
deep-inelastic lepton scattering off nuclei
is considered within the spectator mechanism and a realistic approach in which the rescattering in the medium
of both the
recoiling  proton and the hadronizing nucleon debris  are taken into account.
\end{abstract}
\section{Introduction}
\n Semi inclusive Deep Inelastic Scattering (SIDIS) of leptons
off nuclei
 can provide relevant
information on: (i) possible modification of the nucleon structure function in
 medium (EMC-like effects),
  (ii) the relevance of exotic configurations at short NN distances;
  (iii) the mechanism of quark hadronization
  (see e.g. \cite{FS}).
   A process which attracted much interest from both the theoretical \cite{review,ciosim2}
  and experimental \cite{experiment}
  points of view is the production of slow
  protons,  i.e. the process  A({\cal l},{\cal l}'p)X. In plane wave impulse approximation (PWIA),
  after the hard collision of $\gamma^\star$ with
  a quark of a bound nucleon, two main mechanisms of production of slow protons have been considered;
  they are
depicted in Figs. 1a and 1b and represent, respectively,
the target fragmentation mechanism and
   the spectator mechanism. The latter occurs because of nucleon-nucleon (NN) correlations, namely
    $\gamma^\star$ is absorbed by a nucleon of a correlated
   pair and the second nucleon
   (the spectator
   nucleon) is emitted by recoil and detected in coincidence with the scattered lepton \cite{FS}.
   The spectator mechanism has been intensively investigated (see e.g. \cite{review,ciosim2}),
 but most calculations either completely disregard  the effects of the final state
    interaction,  or
  considered them within simple models. In this contribution
  the first results of the calculation
  of the cross section for the spectator mechanism within a realistic approach
  for the treatment of the FSI are presented. A full account of our approach will be  given elsewhere
  \cite{alciopa}.
\section{The spectator mechanism  within the PWIA}
\noindent In the Bjorken limit  the PWIA cross section for the  spectator mechanism
 is governed, apart from trivial kinematical factors,  by the nuclear structure function
\beqy
F_2^A(x,\Vec{p}_2)&=&\int_x^{M_A/m_N-z_2}
\hspace{-0.5cm}dz_1\,z_1\,F_2^{N}(\frac{x}{z_1})
\int d\Vec{k}_{cm}\,dE^{(2)}\,S(\Vec{k}_{cm}-\Vec{k}_2,\Vec{k}_2,E^{(2)})\times\nonumber\\
&\times&\delta(M_A-m_N(z_1+z_2)-M_{A-2}^fz_{A-2})
\label{qualesara}
\eeqy
where the $z$-axis is directed along the momentum transfer $\bf q$, $\Vec{k}_{cm}=\Vec{k}_1 + \Vec{k}_2=-{\bf P}_{A-2}$,
$F_2^N(\frac{x}{z_1})$ is the structure function of
the hit nucleon, $x=Q/2m_N\nu$ is
the Bjorken scaling variable,
$z_1=k_1^+/m_N$, $z_2=[(m_N^2+{\bf p}_2^2)^{1/2}  - |{\bf p}_2|\cos\theta_2]/m_N$, and
$z_{A-2}=[( (M_{A-2}^f)^2+{\bf k}_{cm}^2 )^{1/2} + ({\bf k}_{cm})_{\parallel}^2]/M_{A-2}^f$ are
 the light-cone momenta of hit nucleon, the detected nucleon and the recoiling spectator nucleus $A-2$,
 respectively,
 and, eventually, $S(\Vec{k}_{1}=\Vec{k}_{cm}-\Vec{k}_{2},\Vec{k}_{2},E^{(2)})$
 is the two nucleon spectral function, with $E^{(2)}$  being the two nucleon removal energy.
 \begin{figure}[!h]
\begin{center}
\includegraphics[width=4cm,height=3.9cm]{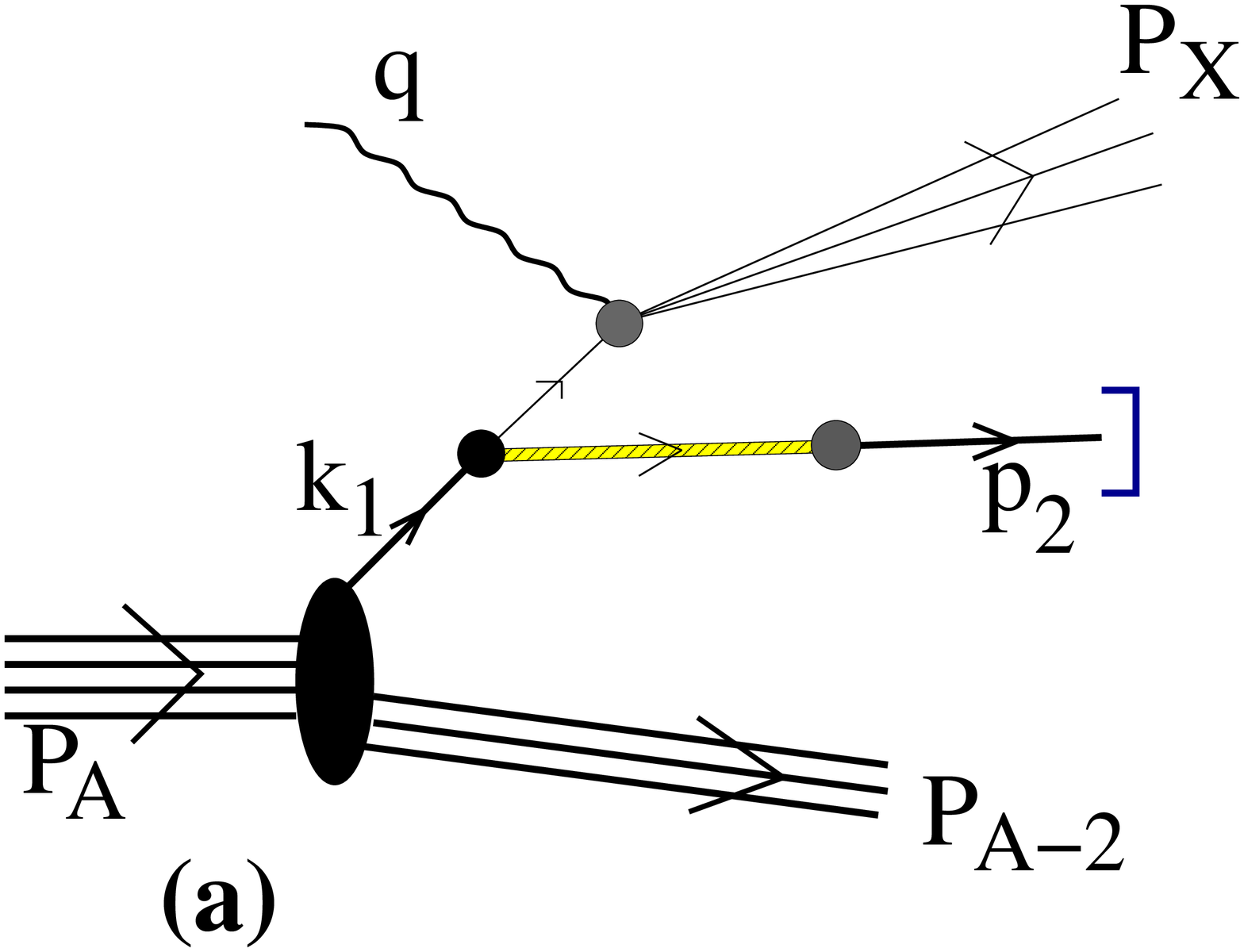}
\hspace{0.7cm}
\includegraphics[width=4cm,height=3.7cm]{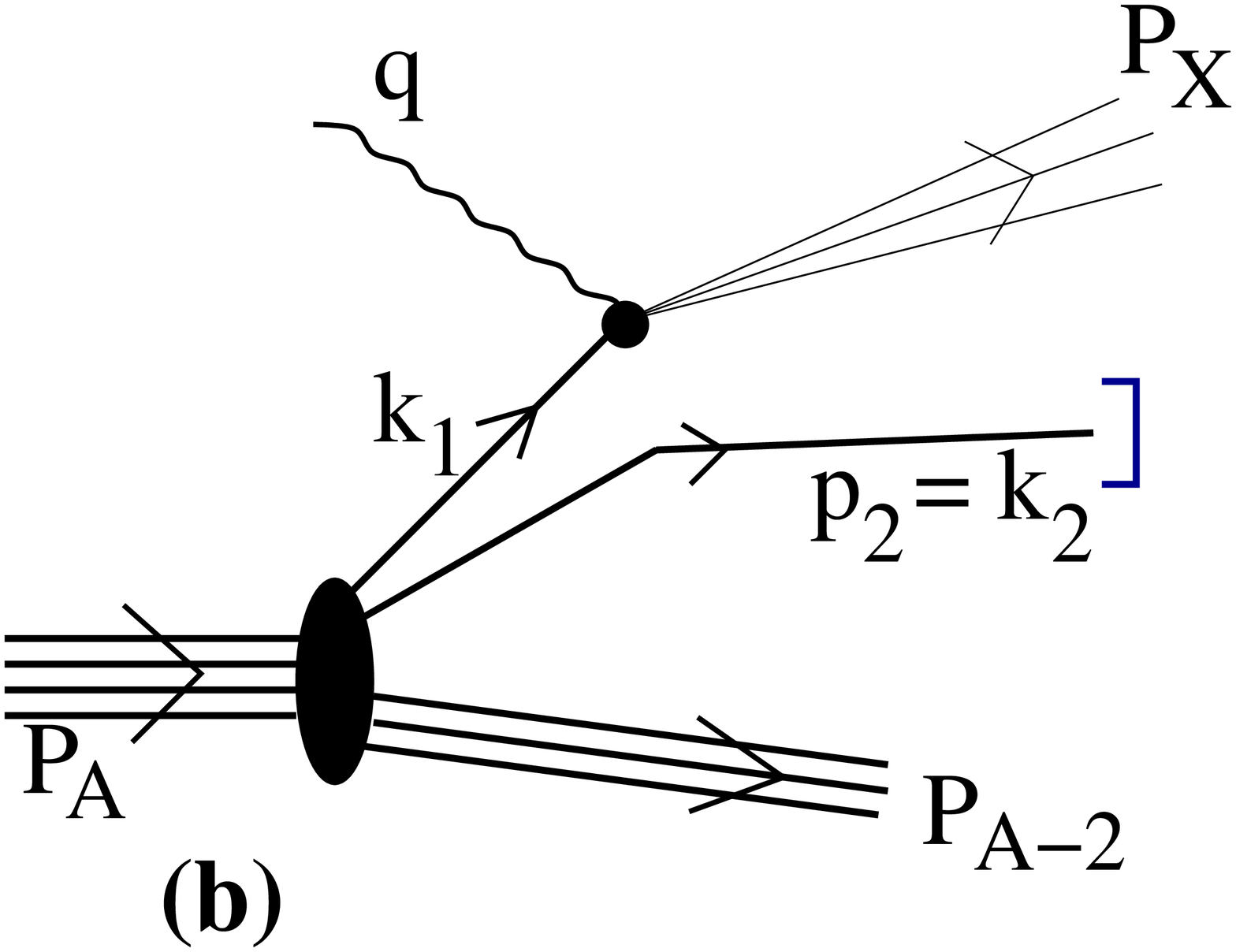}
\hspace{0.7cm}
\includegraphics[width=4cm,height=3.5cm]{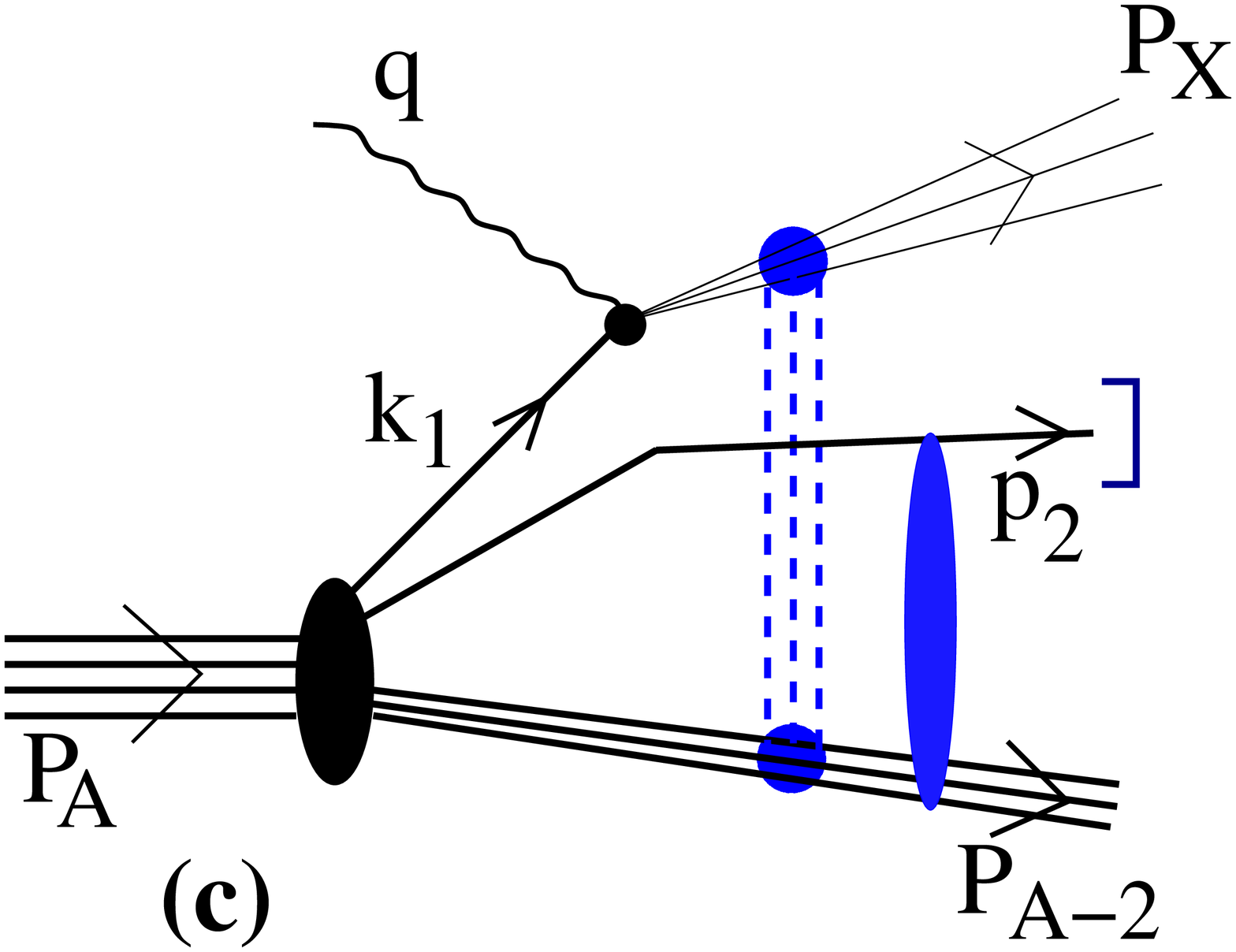}
\vskip -0.1cm
\caption{Slow nucleon production in the process $A(e,e'p)X$ by target fragmentation (a) and by
the spectator mechanism (b). The final state
interaction in the spectator mechanism considered in our approach is depicted in  (c).}
\label{Fig1}
\end{center}
\end{figure}
\vskip -1.5cm
\n Two model spectral functions  have been considered in the literature: i) the two nucleon correlation (2NC)
model \cite{FS}, and ii) the extended 2NC model \cite{ciosim}.
Both of them assume that the momentum and removal energy dependencies  are  only
governed by the dynamics of a
two nucleon correlated pair,  with the $(A-2)$ nucleus acting as a spectator
in the ground state.
In the extended 2NC model the CM motion of the
pair is taken into account and the spectral function
has the form
\beq
\label{2ncex}
S(\Vec{k}_{1},\Vec{k}_{2},E)\,=\,n_{rel}(|\Vec{k}_{1}-\Vec{k}_{2}|/2)
\,n_{cm}(|\Vec{k}_{1}+\Vec{k}_{2}|)\,\delta(E-E_{th}^{(2)})
\eeq
\noindent where $E_{th}^{(2)}$ is the two nucleon emission threshold.
In the  2NC model the correlated pair
is at rest, i.e.  $n_{cm}(\Vec{k}_{cm})=\delta(\Vec{k}_{cm})$;
in this approximation, nuclear effects amount to a momentum-dependent rescaling of the
argument of  $F_{2}^{N}$, namely  $F_{2}^{N}(x/{z}_1)\rightarrow F_{2}^{N}(x/\overline{z}_1)$
 with $\overline{z}_1 \simeq 2-z_2$ \cite{FS};
such a relation is modified by the CM motion of the pair \cite{ciosim2}.
In the present  contribution we go beyond the  PWIA and consider two different kinds of FSI: i) the one of
the nucleon debris  formed after $\gamma^\star$ absorption, with the spectator $A-2$ system; ii) the
one of  the recoiling nucleon  with $A-2$. Both processes, depicted in Fig. 1c,  should strongly affect
 the survival probability
of the
the spectator $A-2$ and, consequently, the cross section of the process.

\section{The final state interaction and the spectator mechanism}

\n Using  momentum conservation
$\Vec{p}_{X}=\Vec{q}-\Vec{p}_{2}-\Vec{P}_{A-2}$,  the transition matrix element of the process $A(e,e'p)(A-2)$
is as follows
\beq
T_{fi}\propto \int \prod_{i=1}^{A}d\Vec{r}_{i}\,
e^{i(\Vec{P}_{A-2}+\Vec{p}_2)\cdot\Vec{r}_1}\,e^{-i\Vec{p}_{2}\cdot\Vec{r}_{2}}\,
\Psi_{A-2}^{\star f}\,\hat{S}_{FSI}^{+}\,\Psi_{A}^{0}\,.
\eeq
Here $\hat{S}_{FSI}$ is the FSI operator, assumed to have the following form

\beq\label{fsiop}
\hat{S}_{FSI}(\Vec{r}_{1},\Vec{r}_{2},\dots,\Vec{r}_{A})\,=\,D_{{\bf p}_2}(\Vec{r}_{2})
\,\prod_{i=3}^{A}G(1,i)
\eeq
where $D_{{\bf p}_2}$ and $\prod_{i=3}^{A}G(1,i)$ take into account the interaction with $A-2$
 of the slow recoiling proton  and
  the fast hit-nucleon-debris, respectively. The first
   is described by an eikonal wave distorted by a complex optical potential, {\it viz}

\beq\label{ottpot}
D_{{\bf p}_{2}}(\Vec{r}_{2})\,=\,exp\left(i\,\frac{E}{p_{2}}\,
\int_{z_{2}}^{\infty}dz\,V(x_{2},y_{2},z)\right)
\eeq

\noindent with the imaginary part of $V(\Vec{r})$ reducing the proton flux. As for the interaction
of the nucleon debris, following \cite{ciokop} we write
$G(1,i)\,=\,1-\theta(z_{i}-z_{1})\,\Gamma(\Vec{b}_{1}-\Vec{b}_{i},z_{i}-z_{1})$, with the z-dependent profile
given by
\beq
\Gamma(\Vec{b}_{1}-\Vec{b}_{i},z_{i}-z_{1})\,=\,\frac{(1-i\alpha)
  \,\sigma_{\mathbf{eff}}(z_{i}-z_{1})}
{4\,\pi b_0 ^2}\,exp\left({-\frac{(\Vec{b}_{1}-\Vec{b}_{i})^{2}}{2\,b_0 ^2}}\right)
\eeq

\noindent where
\begin{figure}[!h]
\begin{center}
\includegraphics[width=7.5cm,height=7.0cm]{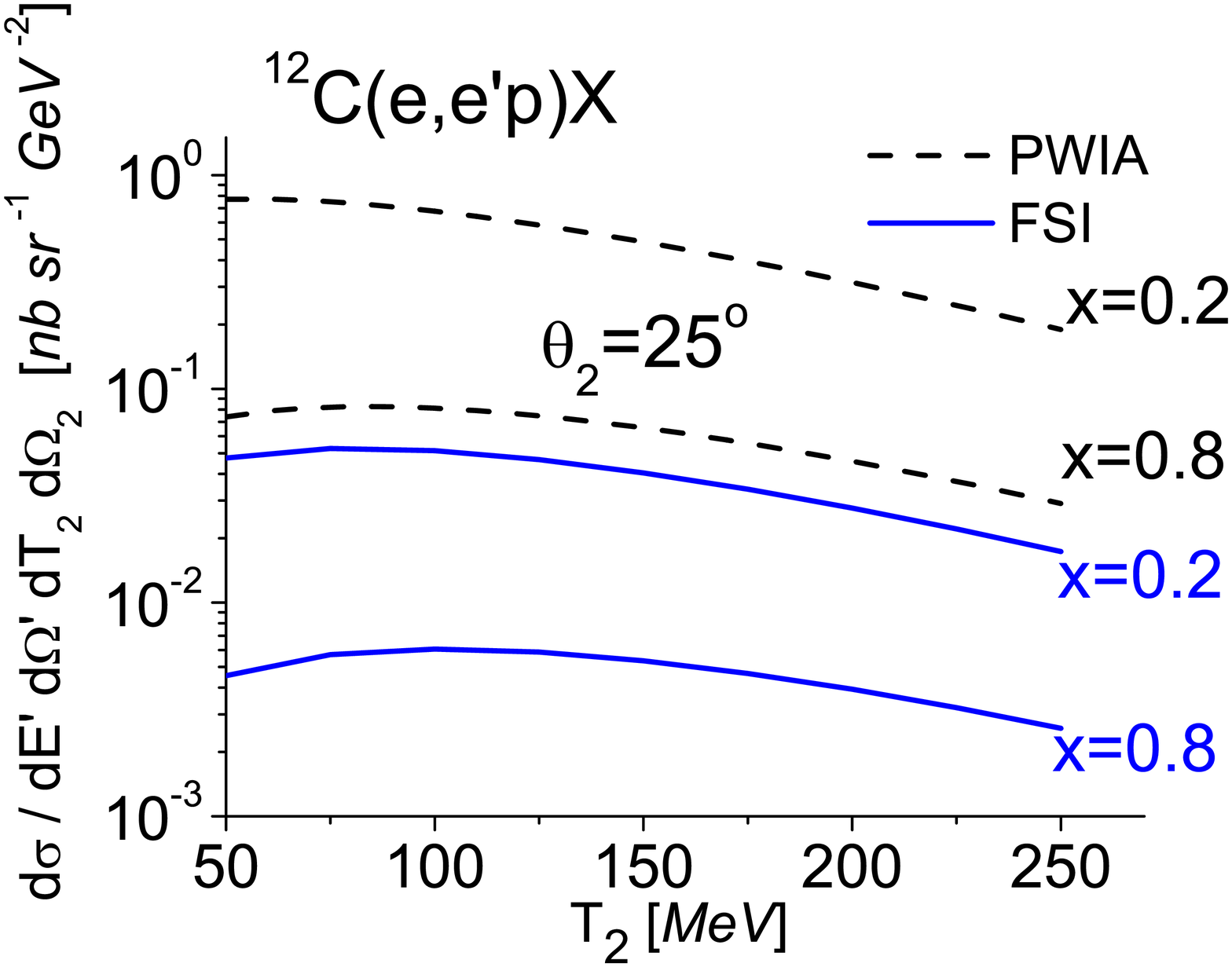}
\includegraphics[width=7.5cm,height=7.0cm]{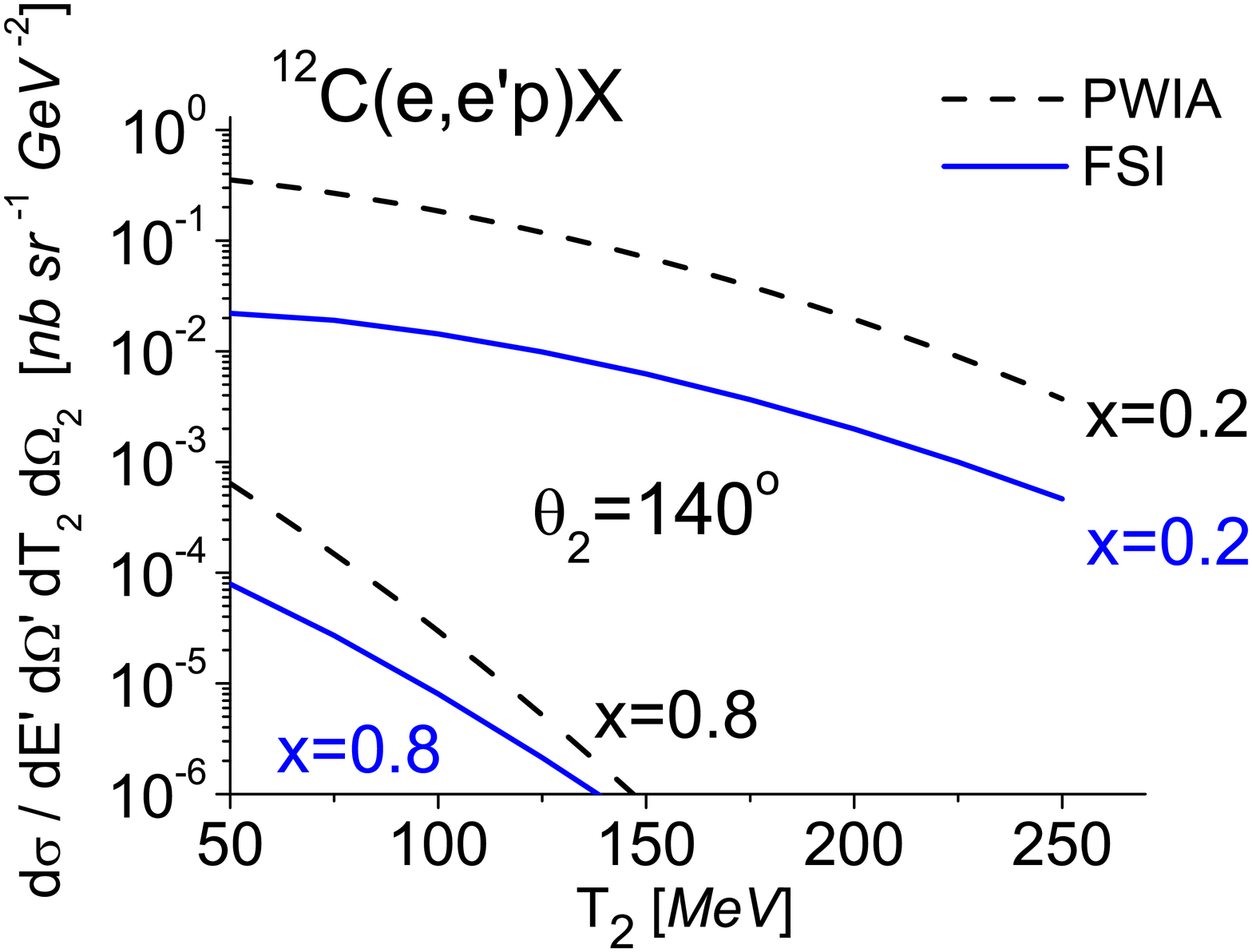}
\vskip -0.5cm
\caption{The DIS differential cross section for the process $^{12}C(e,e'p)X$ versus the kinetic
energy $T_2$ of the detected proton, emitted forward at $\theta_2=25^0$ (\textit{left})
and backward at $\theta_2=140^0$ (\textit{right}), for various values of the Bjorken variable
$x$.
The dashed and full lines represent the results obtained within the extended 2NC model (PWIA)
and the full FSI, respectively.}\label{Fig2}
\end{center}
\vskip -0.7cm
\end{figure}
the novel and main ingredient in our approach  is the time(space)-dependent debris-nucleon cross section
which  has the following form
$\sigma_{eff}(t)\,=\,\sigma_{tot}^{NN}\,+\,\sigma_{tot}^{\pi N}\,
\big[\,n_{M}(t)\,+\,n_{G}(t)\,\big]$; here
the first term describes the interaction of the baryon resulting from
the initial hard $\gamma^\star$- nucleon collision,
whereas the second term represents  the production of (pre)-hadrons (colorless $q\, \overline{q}$ dipoles)
by  the string decay  ($n_{M}(t)$)  and by gluon radiation ($n_{G}(t)$). It has been shown in
\cite{ciokop1} that $\sigma_{eff}(t)$ provides a nice description of grey track production in DIS off nuclei.
The interaction between the debris and the recoiling nucleon, which has been shown to slightly
affect backward nucleon emission at least in the process $^2H(e,e'p)X$
\cite{ciokapko}, will be treated elsewhere \cite{alciopa}.

\section{Results and conclusions}
\noindent We have calculated the differential cross section for the
process $A(e,e^\prime p)(A-2)$ for various nuclei within the PWIA and taking
FSI into account \cite{alciopa}. For the sake of illustration of the effects of FSI,
we show in Fig. 2 the results for the process $^{12}C(e,e'p)^{10}B$ with detection of the
slow proton both in the  forward
($\theta_2=25^o$, $z_2<1$)  and backward  ($\theta_2=140^o$, $z_2>1$) hemispheres
(note that at $x>1$ nucleons from the spectator
mechanism can only be emitted forward). Two main conclusions are worth being mentioned:
i) Backward emission, which can  experimentally be accessed
\cite{experiment},  is very sensitive to nuclear effects
(the 2NC model exhibits kinematical restrictions which are not present in the
extended 2NC model); ii)  FSI appreciably decrease
the cross section, the largest effect arising from the rescattering of the produced (pre)-hadrons
on the nucleon of the spectator $A-2$ system. The approach we have developed  will allow us   to perform
quantitative  comparisons  with available experimental data \cite{experiment} and to provide a
significant answer to
the question as to whether reliable  information on the structure function
of bound nucleons and,  possibly, on hadronization mechanisms could be obtained by means of SIDIS off
 complex nuclei with
slow nucleon production.


%

\begin{thebibliography}{12}
%
\bibitem{FS} L. L. Frankfurt and M. I. Strikman, Phys. Rep. 76 (1981) 216;
  160 (1988) 235
%
\bibitem{review} W. Melnitchouk, M. Sargsian and  M. I. Strikman, Z. Phys. A359 (1997) 99;
G. D. Bosveld, A. E. L. Dieperink, O Scholten, Phys. Rev. C54 (1989) 79;
  S. Simula, Phys. Lett. B387 (1996) 245;
%
C. Ciofi degli Atti, L. P. Kaptari and S. Scopetta,
Eur. Phys. J. A5 (1999) 191.
%
\bibitem{ciosim2} C. Ciofi degli Atti and S. Simula,
  Phys. Lett. B319 (1993) 23
%
\bibitem{experiment} E. Matsinos, et al., Z. Phys. C44 (1989) 79\\
  T. Kitagaki, et al., Phys. Lett. B214 (1988) 281; G. Guy et al,
Phys. Lett. B229 (1989) 421; M. R. Adams et al, Phys. Lett. B319 (1993) 23.
%
\bibitem{alciopa} M. Alvioli, C. Ciofi degli Atti and  V. Palli, to be published
%
\bibitem{ciosim} C. Ciofi degli Atti and S. Simula, Phys. Rev. C53 (1996) 1689.
%
\bibitem{ciokop} C.Ciofi degli Atti and B.Kopeliovich,
  Eur. Phys.J. A17 (2003) 133.
%
\bibitem{ciokop1} C.Ciofi degli Atti and B.Kopeliovich,
 Phys. Lett. B606 (2005) 281.
%
\bibitem{ciokapko} C. Ciofi degli Atti, L.P. Kaptari and B. Kopeliovich,
Eur. Phys. J. A19 (2004) 145
%
\end{thebibliography}
\end{document}